\documentclass[aps,prl,preprint,groupedaddress]{revtex4}

\usepackage{epsfig}
\usepackage{amssymb}
\usepackage{times}

\newcommand{\re}{\mathop{\mathrm{Re}}}

\begin{document}



\title{Optimization of a high efficiency FEL amplifier\footnote{Preprint DESY 14-161, September 2014}}



\author{E.A.~Schneidmiller}
\author{M.V.~Yurkov}
\email[]{mikhail.yurkov@desy.de}
\affiliation{Deutsches Elektronen-Synchrotron (DESY),
Notkestrasse 85, D-22607 Hamburg, Germany}

\date{\today}

\begin{abstract}

The problem of an efficiency increase of an FEL amplifier is now of great
practical importance. Technique of undulator tapering in the post-saturation
regime is used at the existing x-ray FELs LCLS and SACLA \cite{lcls,sacla}, and
is planned for use at the European XFEL, Swiss FEL, and PAL XFEL
\cite{euro-xfel-tdr,swiss-fel,pal-xfel}. There are also discussions on the
future of high peak and average power FELs for scientific and industrial
applications. In this paper we perform detailed analysis of the tapering
strategies for high power seeded FEL amplifiers. Application of similarity
techniques allows us to derive universal law of the undulator tapering.

\end{abstract}


\maketitle


\section{Introduction}

Efficiency of FEL amplifier with untapered undulator is defined by the value of
the FEL parameter $\rho $. Application of the undulator tapering \cite{kroll-tapering}
allows to increase conversion efficiency to rather high values. In the
framework of the one-dimensional theory the status of the problem of tapering
has been settled, and it is generally accepted that optimum law of the
undulator tapering is quadratic with the linear correction for
optimization of the particle's capture in the decelerating potential
\cite{paladin-tap,fawley-scharl,fawley-2lg,fawley-vinokur,ssy-1993,handbook-91,physrep-95,book}.
Similar physical situation occurs in the FEL amplifier with waveguide with
small waveguide parameter. In this case radiation is confined with the
waveguide. Physical parameters of FEL amplifiers operating in infrared,
visible, and x-ray wavelength range are such that these devices are described
in the framework of three dimensional theory with an ``open'' electron beam,
i.e. physical case of pure diffraction in a free space. In this case
diffraction of the radiation is essential physical effect influencing
optimization of the tapering process. Discussions and studies on optimum law of
the undulator tapering in 3D case are in the progress for more than 20 years.
Our previous studies were mainly driven by occasional calculations of
perspective FEL systems for high power scientific (for instance, FEL based
$\gamma \gamma $ - collider ) and industrial applications (for instance, for
isotope separation, and lithography \cite{gg-1995,mw-fel-nim,litho-kw-jm3}).
Their parameter range corresponded to the limit of thin electron beam (small
value of the diffraction parameter). In this case linear undulator tapering
works well from almost the very beginning \cite{ssy-1993}. Comprehensive study
devoted to the global optimization of tapered FEL amplifier with ``open''
electron beam has been presented in \cite{fawley-2lg}. It has been shown
that tapering law should be linear for the case of thin electron beam, optimum
tapering at the initial stage should follow quadratic dependence, and tapering
should start approximately two field gain length before saturation. New wave of
interest to the undulator tapering came with x-ray free electron lasers. It is
used now not only as demonstration tool \cite{bnl-tapering}, but as a routine
tool at operating x-ray FEL facilities LCLS and SACLA \cite{lcls,sacla}.
Practical calculations of specific systems yielded in several empirical laws
using different polynomial dependencies, application of tricks with detuning
jumps, etc (see \cite{wu-tap-2012,gel-tap-2011} and references therein).

In this paper we perform global analysis of the parameter space of seeded FEL
amplifier and derive universal law of the undulator tapering defined by the
only diffraction parameter.

\section{Basic relations}

We consider axisymmetric model of the electron beam. It is assumed that
transverse distribution function of the electron beam is Gaussian, so rms
transverse size of matched beam is $\sigma = \sqrt{\epsilon \beta }$ ,where
$\epsilon = \epsilon_{\mathrm{n}}/\gamma$ is rms beam emittance, $\gamma $ is
relativistic factor, and $\beta $ is focusing beta-function. In the following
we consider rectified case of the ``cold'' electron beam and neglect space
charge effects. Under this assumptions the FEL amplifier is described by the
 diffraction parameter $B$ \cite{book}, and detuning parameter $\hat{C}$:

\begin{equation}
B  = 2 \Gamma \sigma^2 \omega/c \ , \qquad  \hat{C} = C/\Gamma \ ,
\label{eq:reduced-parameters}
\end{equation}

\noindent where $\Gamma = \left[ I \omega^2 \theta_{\mathrm{s}}^2
A_{\mathrm{JJ}}^2/ (I_{\mathrm{A}} c^2 \gamma_{\mathrm{z}}^2 \gamma )
\right]^{1/2}$ is the gain parameter, $C = 2\pi/\lambda _{\mathrm{w}} - \omega
/(2c\gamma ^{2}_{z})$ is the detuning of the electron with the nominal energy
${\cal E}_{0}$. In the following electron energy is normalized as $\hat{P}
= (E-E_0)/(\rho E_0)$, where $\rho = c\gamma ^{2}_{\mathrm{z}}\Gamma /\omega $
is the efficiency parameter (note that it differs from 1-D definition
by the factor $B^{1/3}$ \cite{book}). The following notations are used here:
$I$ is the beam current, $\omega = 2\pi c/\lambda$ is the frequency of the
electromagnetic wave, $\theta_{\mathrm{s}}=K_{\mathrm{rms}}/\gamma$,
$K_{\mathrm{rms}}$ is the rms undulator parameter, $\gamma ^{-2}_{z} = \gamma
^{-2}+ \theta ^{2}_{\mathrm{s}}$, $k_{\mathrm{w}} = 2\pi /\lambda
_{\mathrm{w}}$ is the undulator wavenumber, $I_{\mathrm{A}} =$ 17 kA is the
Alfven current, $A_{\mathrm{JJ}} = 1$ for helical undulator and
$A_{\mathrm{JJ}} = J_0(K_{\mathrm{rms}}^2/2(1+K_{\mathrm{rms}}^2)) -
J_1(K_{\mathrm{rms}}^2/2(1+K_{\mathrm{rms}}^2))$ for planar undulator.  Here
$J_0$ and $J_1$ are the Bessel functions of the first kind.

Equations, describing motion of the particles in the ponderomotive potential
well of electromagnetic wave and undulator get simple form when written down in
normalized form (see, e.g. \cite{book}):

\begin{equation}
\frac{d \Psi}{d \hat{z}} = \hat{C} + \hat{P} , \qquad
\frac{d \hat{P}}{d \hat{z}} = U\cos (\phi_{U} + \Psi) \ ,
\label{eq:pendulum}
\end{equation}

\noindent where $\hat{z} = \Gamma z$, and $U$ and $\phi_{U}$ are amplitude and
phase of effective potential. Energy change of electrons is small in the
exponential stage of amplification, $\hat{P} \ll 1$, and process of electron
bunching in phase $\Psi $ lasts for long distance, $\hat{z} \gg 1$. Situation
changes drastically when electron energy change $\hat{P}$ approaches to the
unity. The change of phase on the scale of $\Delta \hat{z} \simeq 1$
becomes to be fast, particles start to slip in phase $\Psi$ which leads to
the debunching of the electron beam modulation, and growth of the radiation
power is saturated. operation. Undulator tapering \cite{kroll-tapering}, i.e.
adjustment of the detuning according to the energy loss of electrons,
$\hat{C}(\hat{z}) = -\hat{P}(\hat{z}) $, allows to keep synchronism of
electrons with electromagnetic wave and increase output power.

\subsection{Radiation of modulated electron beam}

FEL radiation is coherent radiation of the electron beam which is modulated at
the resonance wavelength during amplification process. It is reasonable here to
remember properties of the radiation of the modulated electron beam. Radiation
power of modulated beam in helical undulator is given by \cite{mod-beam-2005}:

\begin{equation}
P = \frac{\pi\theta^{2}_{\mathrm{s}}\omega I_{0}^{2}a^{2}_{\mathrm{in}}
z}
{4\pi c^{2}}
\left[\arctan\left(\frac{1}{2N}\right)
+ N\ln\left(\frac{4N^{2}}{4N^{2}+1}\right)\right] \ ,
\label{eq:modpow}
\end{equation}

\noindent where $a_{\mathrm{in}}$ is amplitude of modulation of the electron
beam current ($I(z,t) = I_{0} [1 + a_{\mathrm{in}}\cos\omega(z/v_{z}-t)]$), and
$N = k\sigma^{2}/z$ is Fresnel number. We note here that expression
(\ref{eq:modpow}) is a crucial element for understanding the optimum law of
the undulator tapering. Indeed, in the deep tapering regime some fraction of the
particles is trapped in the regime of coherent deceleration. Thus, beam
modulation is fixed, and asymptotically radiation power should be described by
(\ref{eq:modpow}). One can easily find that both asymptotes of undulator
tapering discussed in the introductory section: 1D model of (wide electron
beam), and thin beam asymptote are well described by this expression. Asymptote
of wide electron beam corresponds to large values of Fresnel number $N$, and it
follows from (\ref{eq:modpow}) that radiation power scales as $P \propto z^2$.
Asymptote of thin electron beam corresponds to small values of the Fresnel
Number $N$, and radiation power becomes linearly proportional to the undulator
length, $P \propto z$. Undulator tapering should adjust detuning according to
the energy loss by electrons, and we find that tapering law should be
quadratic for the case of wide electron beam, $C \propto -P \propto z^2$,
and linear - for the case of thin electron beam, $C \propto -P \propto z$.

\begin{figure}[b]

\includegraphics[width=0.6\textwidth]{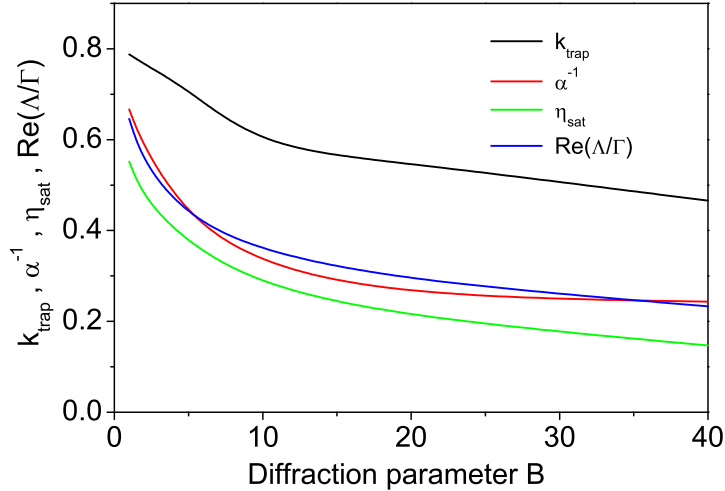}

\caption{
Universal characteristics of FEL amplifier. Color codes are: black - trapping
efficiency $K_{trap}$ for globally optimized undulator; red - fitting
coefficient of global optimization $\alpha _{tap}^{-1}$; blue - FEL field gain
$\re \Lambda /\Gamma $; green - FEL efficiency in the saturation, $\eta_{sat} =
P_{sat}/(\rho P_b)$.}

\label{fig:capt-eff}

\end{figure}

\begin{figure}[b]

\centering
\includegraphics[width=0.6\textwidth]{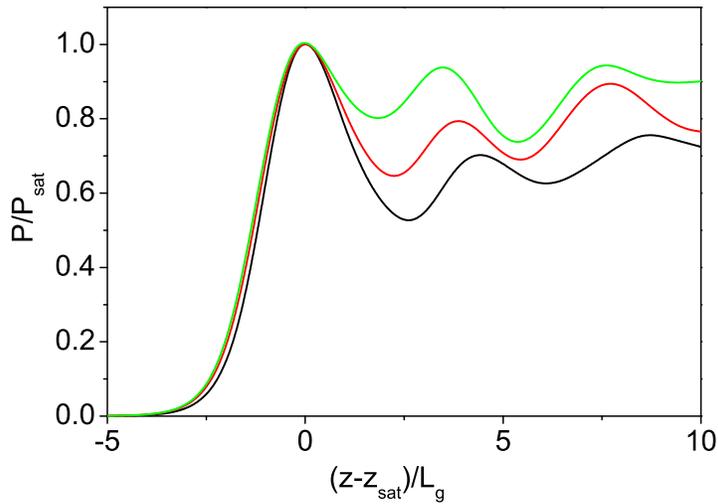}

\caption{
Evolution of the radiation power along the undulator (untapered case).
Color
codes: black, red, green curves correspond to the value of diffraction
parameter $B = 1$, 10, and 40. }

\label{fig:pz-notap}

\end{figure}

\begin{figure}[b]

\centering

\includegraphics[width=0.6\textwidth]{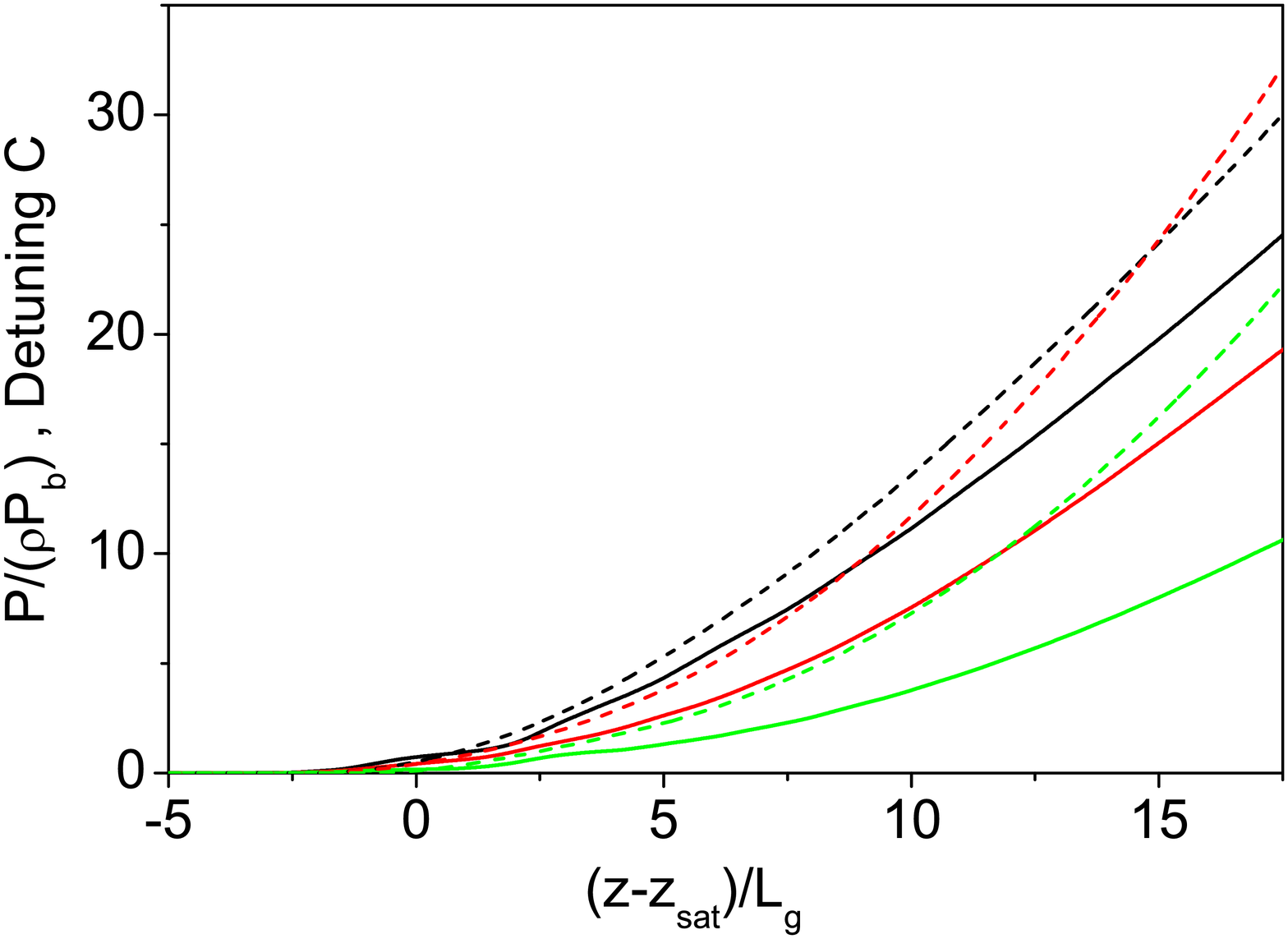}

\caption{
Evolution along the undulator
of the output power (solid curves) and detuning (dashed curves)
for FEL amplifier with global optimization of the undulator
tapering.
Color
codes: black, red, green curves correspond to the value of diffraction
parameter $B = 1$, 10, and 40. }

\label{fig:pz-tap}



\centering
\includegraphics[width=0.6\textwidth]{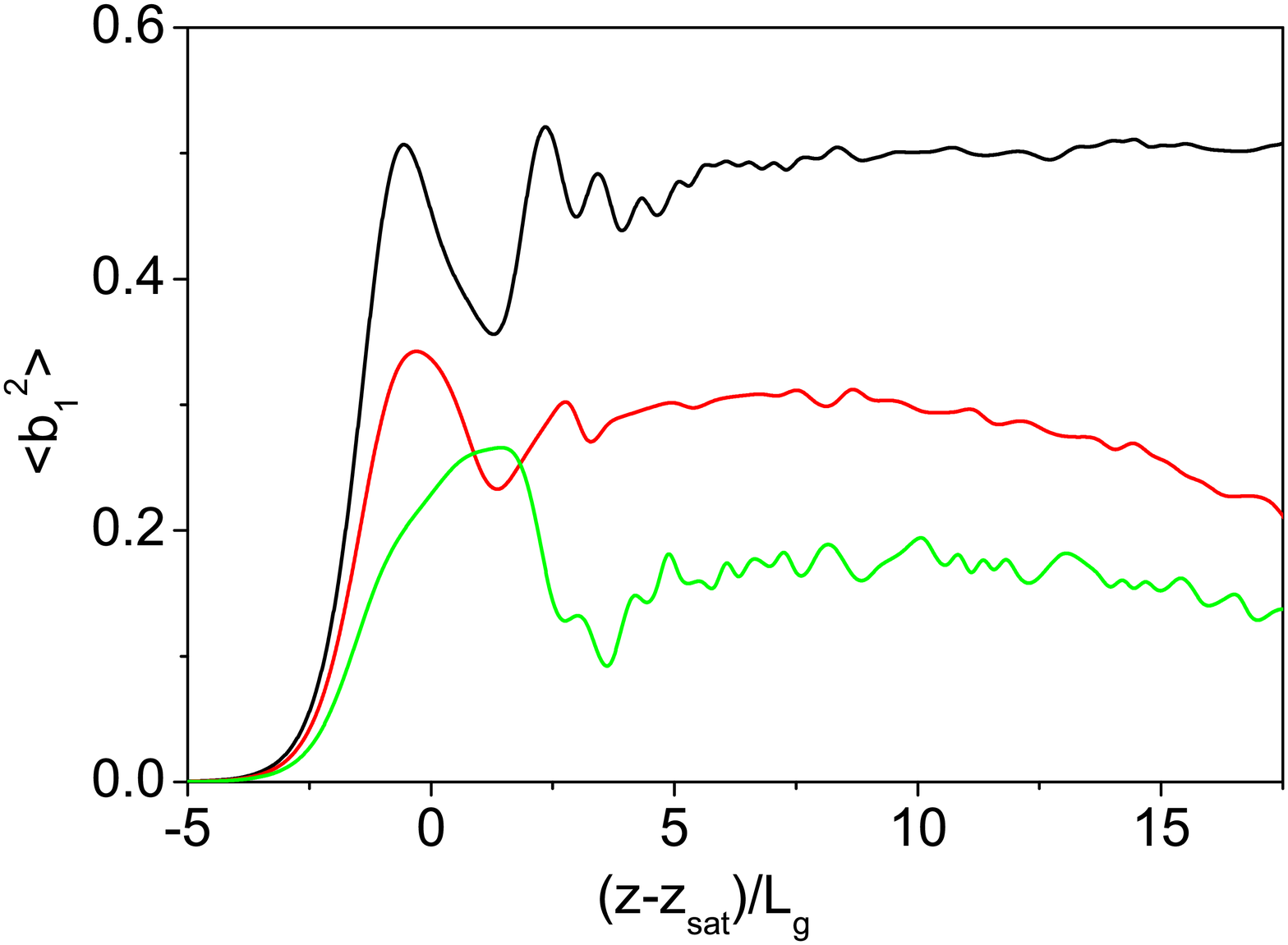}

\caption{
Evolution along the undulator of the squared value of the beam bunching for FEL
amplifier with global undulator tapering. Color codes: black, red, green curves
correspond to the value of diffraction parameter $B = 1$, 10, and 40. }

\label{fig:a2-tap}

\end{figure}

\begin{figure}[tb]

\centering
\includegraphics[width=0.6\textwidth]{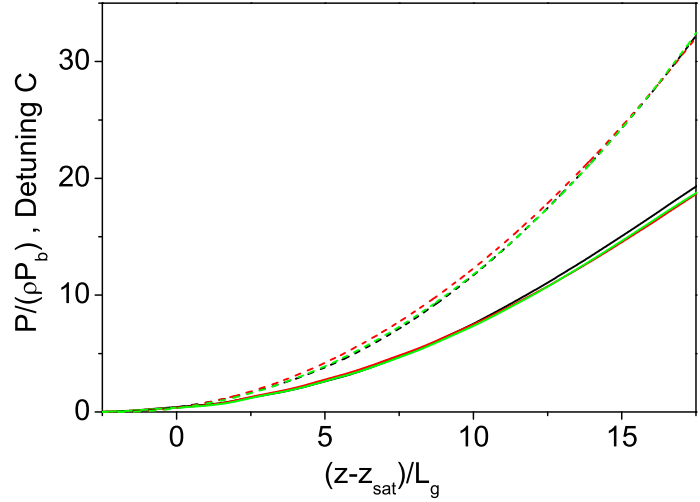}

\caption{
Evolution along the undulator
of the output power (solid curves) and detuning (dashed curves)
Color
codes: black - FEL with global optimization of undulator tapering, red - fit
with formula (\ref{eq:tapfit}), green - fit with rational function
(\ref{eq:ratfit}) Here the value of diffraction parameter is $B = 10$. }

\label{fig:pz-tap-b10}

\end{figure}

\begin{figure}[tb]

\centering
\includegraphics[width=0.6\textwidth]{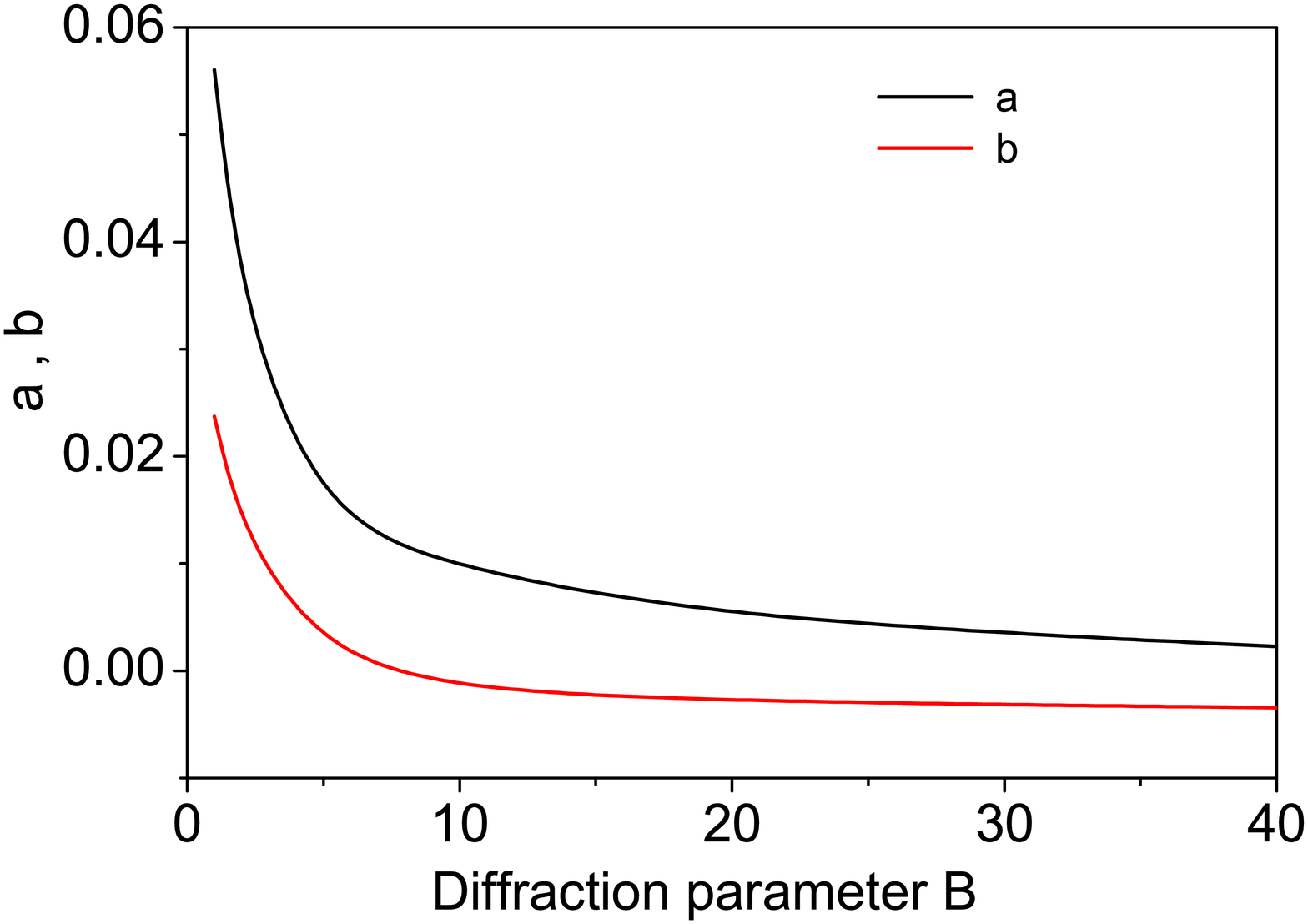}

\caption{
Coefficients $a$ (black line) and $b$ (red line) of the rational fit of the
tapering law (\ref{eq:ratfit}).
}

\label{fig:ratfit}

\end{figure}

\section{Global optimization}

We start with global optimization of the parameter space. Simulations have been
performed with three-dimensional, time-dependent FEL simulation code FAST
\cite{fast}. In the framework of the accepted model (cold electron beam) both,
field gain $\re \Lambda /\Gamma $ and efficiency in the saturation, $\eta_{sat}
= P_{sat}/(\rho P_b)$ of the FEL amplifier tuned to exact resonance are defined
by the only diffraction parameter $B$ (see Fig.\ref{fig:capt-eff}). Operation
of the FEL amplifier before saturation is also defined by the diffraction
parameter $B$. One can clearly observe this from Fig.~\ref{fig:pz-notap}. Here
longitudinal coordinate is normalized to the gain length $L_{g} = 1/(\re\Lambda
/\Gamma )$, and radiation power is normalized to the saturation power. When
amplification process enters nonlinear stage, output power is function of two
parameters, diffraction parameter and reduced undulator length.

Now we come to the problem of efficiency increase with undulator tapering.
First, we solve this problem using approach of straightforward global
optimization. The function of optimization is to find maximum of the output
power at the undulator length exceeding ten field gain lengths. We divide
undulator into many pieces and change detuning of all pieces independently. We
apply adiabatic (smooth) tapering, i.e. we prevent jumps of the detuning on the
boundary of the sections. Number of sections is controlled to be large enough
to provide the result which is independent on the number of sections. Then we
choose tapering law $C(B,z)$ corresponding to the maximum power at the exit of
the whole undulator. This global optimization procedure has been performed in
the practically important range of diffraction parameters from $ B = 1$ to $B =
40$. Results of this global optimization are summarized in Fig.
\ref{fig:pz-tap}. Ratio of the normalized power to the normalized detuning
gives us the value of trapping efficiency of electrons into the regime of
coherent deceleration, $K_{trap} = \hat{P}/\hat{C}$. This universal function of
diffraction parameter $B$ is plotted on Fig:~\ref{fig:capt-eff}. We find that
optimum trapping factor approaches values of 80\% for $B = 1$, and falls down
to 45\% for $B = 40$. It is interesting to notice that for $B \gtrsim 5$ it
scales roughly as $B^{-1/3}$, similar to other FEL characteristics like FEL gain
and saturation efficiency.

\section{Universal tapering law}

It comes from global optimization that in the whole parameter range undulator
tapering starts from the value of $\Delta z \simeq 2 L_g$ before saturation.
This is not surprising if we look on Fig.~\ref{fig:pz-notap}. Optimum undulator
tapering should compensate loss of the electron energy which is in fact follows
identical parametric dependence on the gain $L_g$ for all values of
diffraction parameter. Next observations come from the analysis of the beam
modulation. The first observation is that the beam modulation at the initial
stage of the nonlinear regime follows similar behavior for all diffraction
parameters (see Fig.~\ref{fig:a2-tap}). This gives a hint that initial capture
of the particles is performed in a similar way in the whole parameter range.
The second observation is that the beam modulation after trapping of the
electrons to the coherent deceleration process remains constant along the
undulator, and it is universal function of the diffraction parameter $B$ (see
Fig.~\ref{fig:a2-tap}). This is gives us the main hint which we discussed in
the previous section. I.e., excluding trapping transition stage, we deal with
radiation of the modulated electron beam (\ref{eq:modpow}).  Main essence of our
study is to apply parametrical dependence like (\ref{eq:modpow}) to fit optimum
detuning pattern in Fig.~\ref{fig:pz-tap} such that condition of optimum tapering
is preserved:

\begin{equation}
\hat{C} = \alpha _{tap} (\hat{z}-\hat{z}_0)
\left[\arctan\left(\frac{1}{2N}\right)
+ N\ln\left(\frac{4N^{2}}{4N^{2}+1}\right)\right] \ ,
\label{eq:tapfit}
\end{equation}

\noindent with Fresnel number $N$ fitted by $N = \beta
_{tap} / (\hat{z}-\hat{z}_0)$. Thus, we try to fit optimum detuning with three
parameters: $z_0$, $\alpha _{tap}$ and $\beta _{tap}$. Here undulator length is
normalized to the gain parameter, $\hat{z} = \Gamma z$. One parameter of this
fit, start of the undulator tapering $z_0$ is firmly fixed by the global
optimization procedure, $z_0 = z_{sat}-2L_g$. Another parameter of the problem,
$\beta _{tap}$, is rather well approximated with the linear dependency on
diffraction parameter, $\beta _{tap} = 8.5 \times B$. Remaining parameter,
$\alpha _{tap}$ is plotted in Fig.~\ref{fig:capt-eff}. It is slow function of
the diffraction parameter $B$, and scales approximately to $B^{1/3}$ as all
other important FEL parameters presented in Fig.~\ref{fig:capt-eff}. Thus,
application of similarity techniques gives us an elegant way for general
parametrical fit of such complicated phenomena as optimum undulator tapering.
Actually, accuracy of this fit is pretty good giving the results for optimum
detuning which are close to the global optimum. We illustrate with
Fig.~\ref{fig:pz-tap-b10} tapering law (\ref{eq:tapfit}) for specific value of
the diffraction parameter $B = 10$. Curves in black color are normalized power
and detuning derived from global optimization. Red dashed curve is detuning
$\hat{C}$ given by (\ref{eq:tapfit}) with $\alpha _{tap} = 3.6$ (see
Fig.~\ref{fig:capt-eff}, and $\beta _{tap} = 85$ (according to relation $\beta
_{tap} = 8.5 \times B$). The solid curve in red color is normalized FEL
efficiency simulated using detuning (\ref{eq:tapfit}). We see good agreement of
the fit with global optimization. The same situation occurs in the whole range
of traced values of diffraction parameter $B$. Such a good agreement is not
surprising since fitting is based on very clean parametric dependencies, and
numerical simulations just provided relevant numerical factors.

\subsection{Rational fit}

Analysis of expression (\ref{eq:tapfit}) shows that it has quadratic
dependence in $z$ for small values of $z$ (limit of wide electron beam),
and linear dependence in $z$ for large values of $z$ (limit of thin electron
beam). Natural idea comes to try fit with rational function which satisfies
both asymptotes. The simplest rational fit is:

\begin{equation}
\hat{C} = \frac{a (\hat{z}-\hat{z}_0)^2}
{1+b (\hat{z}-\hat{z}_0)} \ .
\label{eq:ratfit}
\end{equation}

\noindent Coefficients $a$ and $b$ are universal functions of diffraction
parameter $B$, and are plotted in Fig.~\ref{fig:ratfit}.
Start of the undulator tapering is set to the value $z_0 = z_{sat}-2L_g$
suggested by the global optimization procedure.  Analysis of plots presented in
Fig.~\ref{fig:pz-tap-b10} shows that fit of the universal tapering law with
rational also works well.

\subsection{Trapping process}

\begin{figure}[tb]

\centering
\includegraphics[width=65mm]{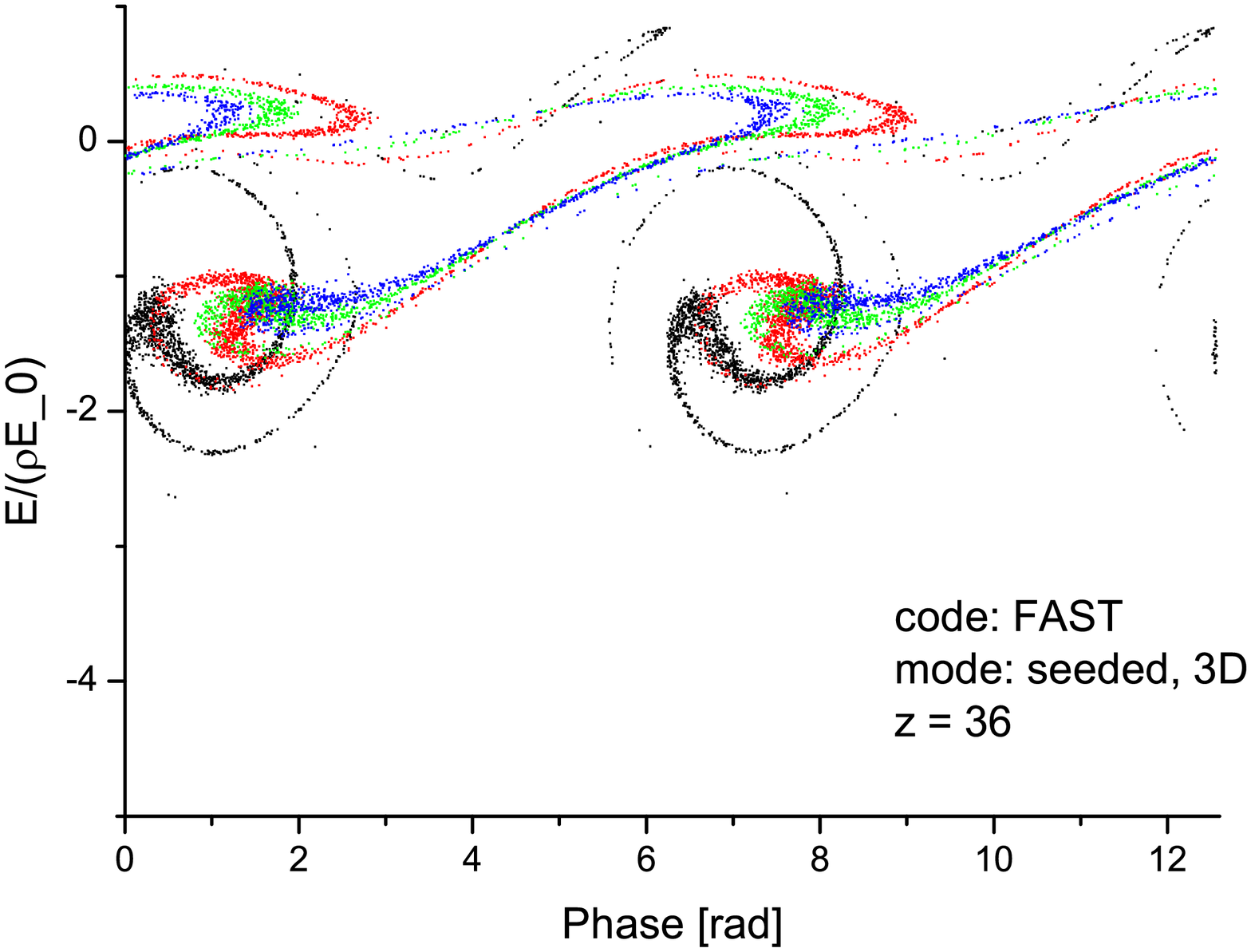}
\includegraphics[width=65mm]{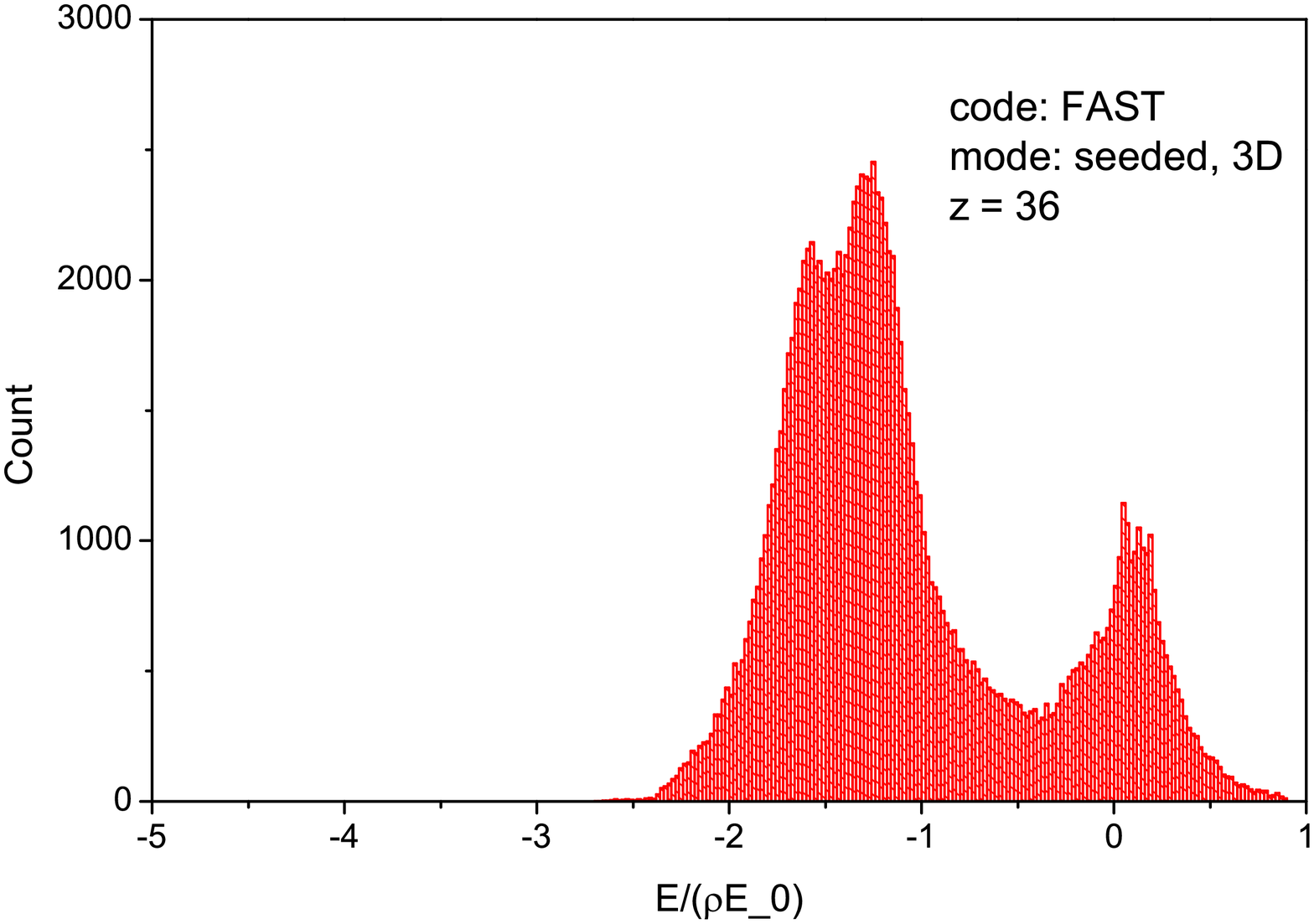}
\includegraphics[width=65mm]{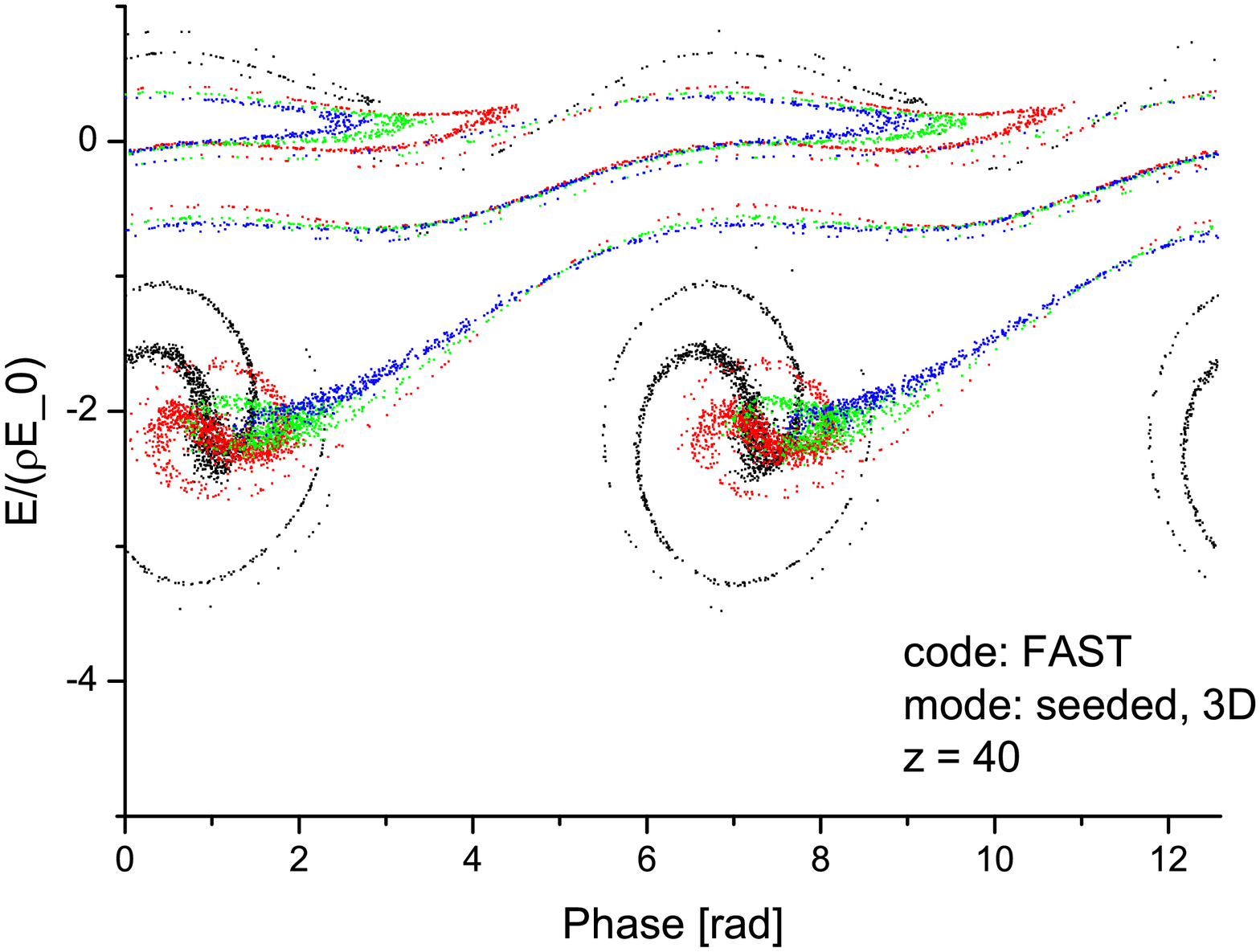}
\includegraphics[width=65mm]{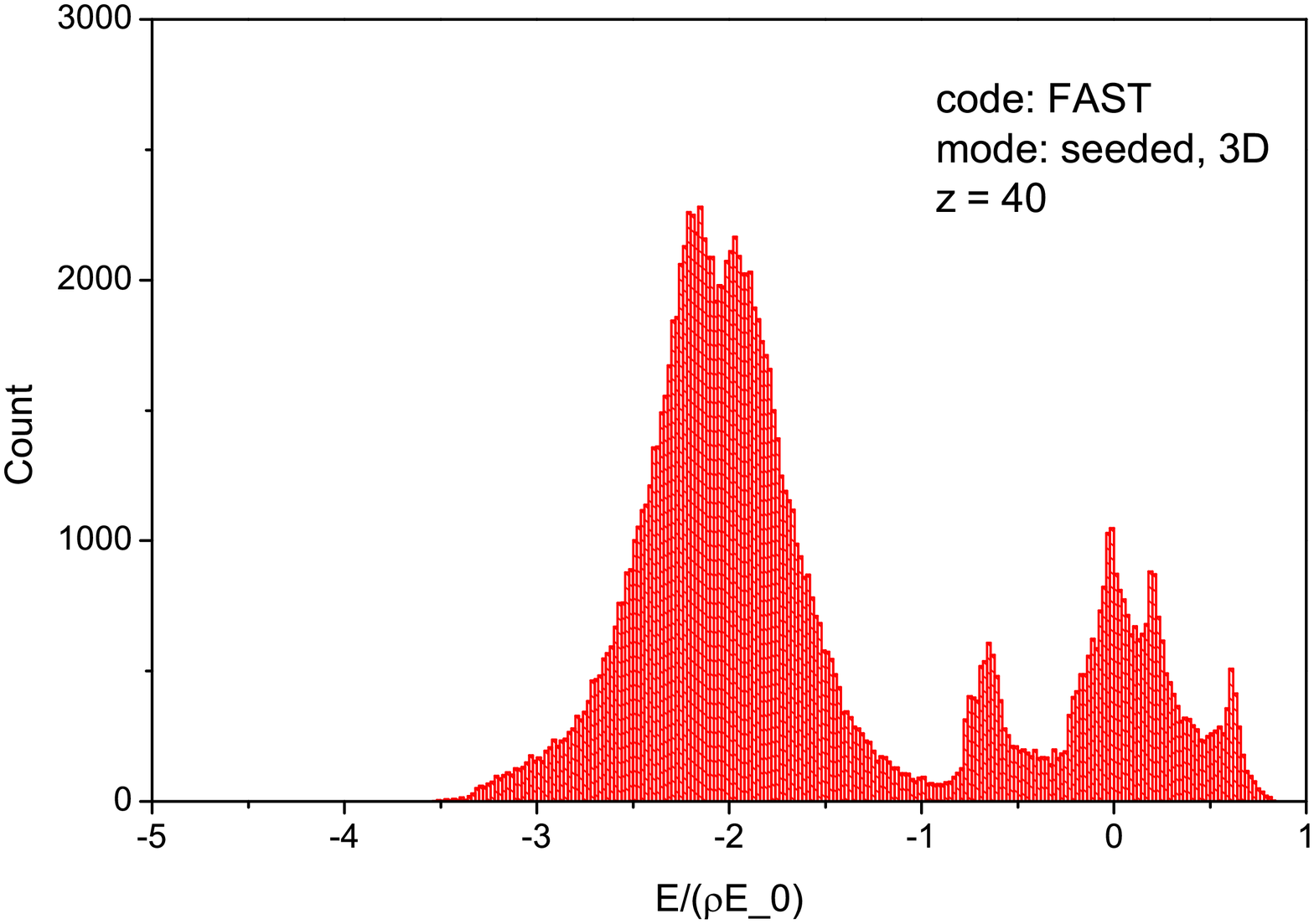}
\includegraphics[width=65mm]{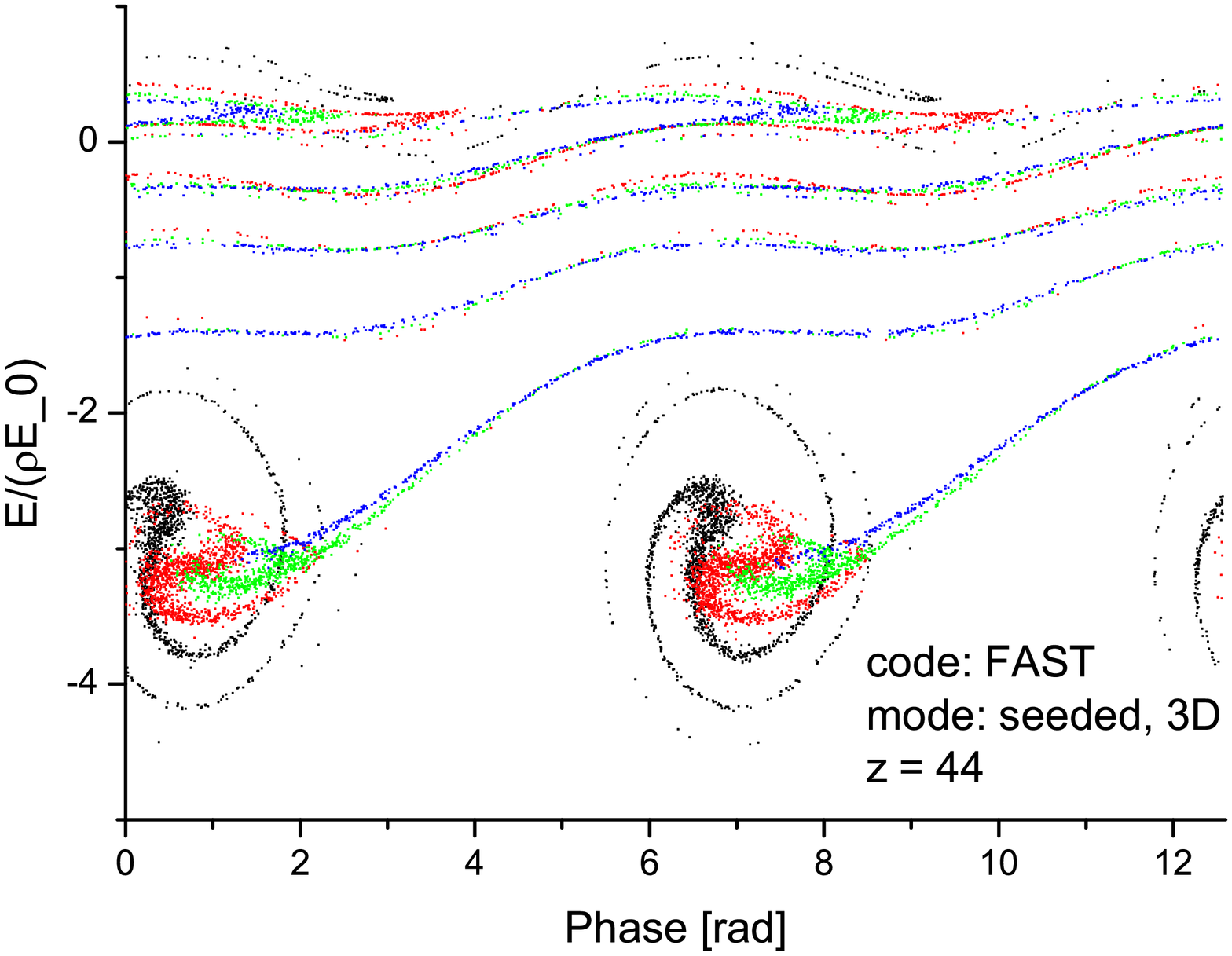}
\includegraphics[width=65mm]{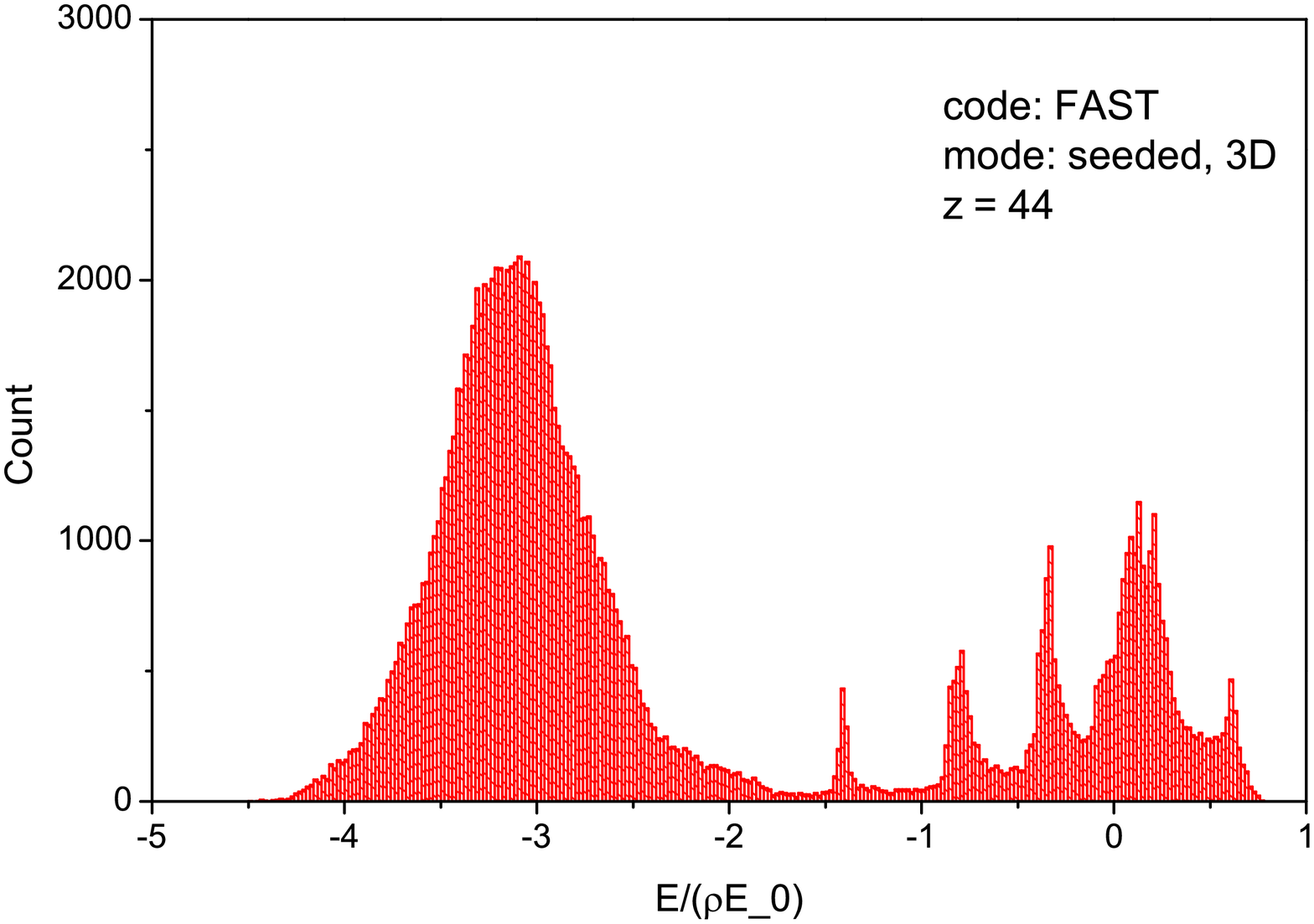}

\caption{
Phase space distribution of electrons
(left column) and Population of electrons in energy (right column)
at different stages of trapping process.
Color codes correspond to different location of the particles in
the beam (black - core of the beam, blue - edge of the beam). Here diffraction
parameter is $B = 10$. Top, middle, and bottom plots correspond to
$(z-z_{sat})/L_g$ = 2.5, 3.9 and 5.3, respectively
(see Fig:~\ref{fig:pz-tap-b10}).
}

\label{fig:ps-b10}

\end{figure}

We finish our paper with illustration of the trapping process. Trapping
efficiency $K_{trap} = \hat{P}/\hat{C}$ is plotted in Fig.~\ref{fig:capt-eff}.
Trapping efficiency falls down with diffraction parameter $B$. This is natural
consequence of diffraction effects discussed earlier (see, e.g. \cite{book},
Chap. 4). For small value of the diffraction parameter $B$ gradient of the
field of the beam radiation mode across the electron bunch is smaller than for
large values of diffraction parameter. In the latter case we obtain situation
when electrons located in the core of the electron beam are already fully
bunched while electrons on the edge of the beam are not bunched yet. As a
result, number of electrons with similar positions on the energy-phase plane
falls down with the growth of the diffraction parameter, as well as trapping
efficiency into the regime of coherent deceleration. The trapping process is
illustrated with phase space plots presented on Fig.~\ref{fig:ps-b10} for the
value of diffraction parameter $B = 10$.
Top, middle, and bottom plots correspond to
the points of $(z-z_{sat})/L_g$ = 2.5, 3.9 and 5.3 on Fig.~\ref{fig:pz-tap}.
Different color codes (black to blue) correspond to
different locations of the particles across the beam (core to edge.
We see that particles in
the core of bunch (black points) are trapped most effectively. Nearly all
particles located in the edge of the electron beam (blue points) leave
stability region very soon. Trapping process lasts for several field gain
length when trapped particles become to be isolated in the trapped energy band
for which undulator tapering is optimized further. For specific value of the
diffraction parameter $B = 10$ it is not finished even at three field gain
lengths after saturation, and non-trapped particles continue to populate low
energy tail of the energy distribution (see right column of
Fig.~\ref{fig:ps-b10}). Recently we have been invited to the discussion on the
details of trapped particles distribution in the phase space observed
experimentally at LCLS \cite{behrens-bands1}. Graphs presented in
Fig.~\ref{fig:ps-b10} give a hint on the origin of energy
bands which are formed by non-trapped particles. This is consequence of
nonlinear dynamics of electrons leaving the region of stability. Actually,
similar effect can be seen in the early 1D studies
\cite{handbook-91,physrep-95}.

\section{Discussion}

In this paper we derived general law for optimum undulator tapering in the
presence of diffraction effects (\ref{eq:tapfit}). Purified case of ``cold''
electron has been considered. This allowed us to isolate diffraction effects in
the most clear form. It has been found that universal function of the undulator
tapering depends on the only diffraction parameter. Fit of the universal
tapering law with rational function (\ref{eq:ratfit}) requires fulfillment of
two asymptotes of the tapering law: quadratic at the initial stage (wide beam
asymptote), and linear for very long tapering section (thin beam asymptote). It
is essentially simple, and can be very convenient for optimization of
practical systems. Tapering law is described with simple analytical expressions
with two fitting coefficients. Extension of this approach to practical life
(including energy spread and emittance) is pretty much straightforward and will
result in corrections to the fitting coefficients without changing general law
given by (\ref{eq:tapfit}). The same law is evidently applicable to SASE FEL as
well with relevant correction of fitting coefficients.

\section{Acknowledgement}

We are grateful to William Fawley for useful discussion on the problem of
undulator tapering which continued through many years. We are grateful to Vadim
Banine and Vivek Bakshi, contacts with them and other members of industrial
community stimulated our interest to the development of high power FEL systems.
We thank Christopher Behrens for attracting our attention to deeper analysis of
the trapping process (energy bands) and for useful discussions.

\end{document}